\documentclass[12pt]{article}
\pdfoutput=1

\usepackage[bulletsep]{collref}
\usepackage{amssymb,graphicx}
\usepackage[intlimits]{amsmath}
\usepackage{bbm}
\usepackage[small]{subfigure}
\usepackage{isak-latex-sty}
\usepackage{multirow}
\usepackage{booktabs}

\usepackage{hyperref}
\usepackage{color}

\usepackage{MnSymbol}

\makeatletter \@addtoreset{equation}{section} \makeatother

\makeatletter
\let\old@startsection=\@startsection
\let\oldl@section=\l@section
\renewcommand{\@startsection}[6]{\old@startsection{#1}{#2}{#3}{#4}{#5}{#6\mathversion{bold}}}
\renewcommand{\l@section}[2]{\oldl@section{\mathversion{bold}#1}{#2}}
\makeatother

\makeatletter
\let\old@makecaption=\@makecaption
\def\@makecaption{\small\old@makecaption}
\makeatother

\renewcommand{\leq}{\leqslant}
\renewcommand{\geq}{\geqslant}

\newcommand{\sfrac}[2]{{\textstyle\frac{#1}{#2}}}
\newcommand{\half}{\sfrac{1}{2}}
\newcommand{\ihalf}{\sfrac{i}{2}}

\newcommand{\neel}{{\rm N\acute{e}el}}
\newcommand{\mps}{{\rm MPS}}
\newcommand{\ur}{\uparrow}
\newcommand{\dr}{\downarrow}

\newcommand{\xx}{\mathtt{x}_1}
\newcommand{\yy}{\mathtt{x}_2}

\begin{document}


\setcounter{page}{1}
\renewcommand{\thefootnote}{\arabic{footnote}}
\setcounter{footnote}{0}

\begin{titlepage}

\begin{flushright}\footnotesize
\texttt{NORDITA-2015-132} \\
\texttt{UUITP-26/15}
\vspace{0.3cm}
\end{flushright}

\centerline{\large \bf One-point Functions in AdS/dCFT from Matrix Product States}
\vskip 0.7 cm

\centerline{{\bf Isak Buhl-Mortensen$\,^{1}$},   {\bf Marius de Leeuw$\,^{1}$},  {\bf Charlotte Kristjansen$\,^{1}$} and {\bf Konstantin Zarembo$\,^{2}$}  }

\vskip 0.6cm

\begin{center}
\sl $^1$ The Niels Bohr Institute, University of Copenhagen \\
\sl  Blegdamsvej 17, DK-2100 Copenhagen \O , Denmark
\vskip 0.3cm
\sl $^2$ NORDITA, KTH Royal Institute of Technology and Stockholm University \\
Roslagstullsbacken 23, SE-106 91 Stockholm, Sweden  \\
Department of Physics and Astronomy, Uppsala University\\
SE-751 08 Uppsala, Sweden
\end{center}
\vskip 0.7cm

\centerline{\small\tt buhlmort@nbi.ku.dk, deleeuwm@nbi.ku.dk, kristjan@nbi.ku.dk, zarembo@nordita.org}

\vskip 1.2cm \centerline{\bf Abstract} \vskip 0.2cm \noindent

\noindent

One-point functions of certain non-protected scalar operators in the defect CFT dual to the D3-D5 probe brane system with $k$ units
of world volume flux can be expressed as overlaps between Bethe eigenstates of the Heisenberg spin chain and  a matrix product state. We
present a closed expression of determinant form for these one-point functions, valid for any value of $k$. 
The determinant formula factorizes into the $k=2$ result times a $k$-dependent pre-factor.  
Making use of the transfer matrix of the Heisenberg spin chain we recursively relate the matrix product state for higher even and odd $k$ to the matrix product state for $k=2$ and $k=3$ respectively. 
We furthermore find evidence that the matrix product states for $k=2$ and $k=3$ are
related via a ratio of Baxter's $Q$-operators.
The general $k$ formula has an interesting thermodynamical limit involving a non-trivial scaling of $k$, which indicates that
the match between string and field theory one-point functions found for chiral primaries might be tested for non-protected operators as well. We revisit  the string computation for chiral primaries and discuss how it can be extended to non-protected operators.

\end{titlepage}


\section{Introduction}
Holographic modeling of spontaneously or explicitly broken symmetries typically involves probe branes.
An interesting class of quantum field theory set-ups arises when the probe brane breaks translational invariance and introduces a defect in the dual field theory. Internal degrees of freedom on the defect then originate from open strings and belong to the fundamental representation of the gauge group, while the fields in the bulk arise from  closed strings and transform in the adjoint.
Such defect field theories allow for novel types of correlation functions that are not possible without the defect. 
Examples are one-point functions of bulk fields and correlation functions involving operators localized on the defect.

In the present paper we concentrate on the defect CFT  dual to the D3-D5 probe brane system with $k$ units of background gauge field 
flux~\cite{Karch:2000gx}. The brane intersection introduces a domain wall that separates the vacua with 
respectively unbroken $SU(N)$ and $SU(N-k)$ gauge symmetry in the $\mathcal{N}=4$ supersymmetric Yang-Mills (SYM) theory, with additional degrees of freedom living on the defect \cite{deWolfe:2001pq,Erdmenger:2002ex}. 
One-point functions in this dCFT were studied in~\cite{deWolfe:2001pq,Nagasaki:2012re,Kristjansen:2012tn,deLeeuw:2015hxa} whereas two-point functions of defect operators were
considered in~\cite{deWolfe:2001pq,DeWolfe:2004zt,Susaki:2004tg,McLoughlin:2005gj,Susaki:2005qn,Okamura:2005cj}, where integrability of the underlying $\mathcal{N}=4$ SYM proved particularly useful. The defect operators are mapped to spin chains with open boundary conditions and are dual to open strings attached to the probe D5-brane.  We approach the problem from a different angle by picturing the D5-brane as a boundary state that can emit and absorb closed strings. An absorption of a single string state is represented, in the field theory, by a one-point function of a bulk operator. 

The non-vanishing flux on the D5-brane represents $k$ D3 banes dissolved in its world-volume, while  in the field-theory language the symmetry-breaking is described by a non-zero vacuum expectation value of scalar fields, that form a $k$-dimensional unitary representation of $\mathfrak{su}(2)$ \cite{Diaconescu:1996rk,Giveon:1998sr,Constable:1999ac}. In the present paper we  continue our study~\cite{deLeeuw:2015hxa} of the one-point functions in the defect CFT resulting from this semiclassical description. We also do some rudimentary analysis on the strong-coupling side of the AdS/CFT duality.  

Our work relies in many ways on methods borrowed from solid state physics.
 It is already well-known that probe brane
systems can be used to model various strongly coupled condensed matter systems (see  \cite{Ammon:2015wua} for an overview). Furthermore,
the spin-chain picture of the single-trace operators in  $\mathcal{N}=4$ SYM
uncovers the integrable structure of the theory~\cite{Minahan:2002ve,Beisert:2010jr} and paves the way for the use of the Bethe ansatz techniques that greatly facilitate the spectral analysis of theory. Apart from these well-known points of
contact we find that so-called matrix product states (MPS), which in the condensed matter context have been used in the evaluation of quantum entanglement in one-dimensional systems, have exactly the right properties to act as a "defect state". The computation of the one-point functions in the dCFT maps to the computation of an overlap between the MPS and the Bethe eigenstates of the spin chain.
Finally, the N\'{e}el state, i.e.\  the ground state of
the classical Heisenberg anti-ferromagnet, plays a surprisingly central r\^ole in our investigations. 

 A simple set of scalar operators in the D3-D5 dCFT  with non-trivial one-point functions are the operators  of the form $\mathop{\mathrm{tr}}Z^{L-M}W^M$, where $Z$ and $W$ are complex scalar fields from the $\mathcal{N}=4$ supermultiplet. Conformal operators  belonging to this $SU(2)$ 
 sub-sector are known to be expressible as Bethe eigenstates of the Heisenberg XXX$_{1/2}$ spin 
 chain of length $L$ in the sector with $L-M$ spins up and $M$ spins down. Each operator is characterized by a set of $M$
 Bethe roots and, as shown in \cite{deLeeuw:2015hxa}, only parity-symmetric operators with paired rapidities $\{u_j, -u_j\}_{j=1}^{M/2}$ and even length, $L$, can have non-trivial one-point functions at tree level. 
 The one-point functions are constrained by conformal symmetry to take the form
 \begin{align}
 \langle {\cal O}_{L} (x)\rangle = \frac{C_k(\{u_j\})}{x^{L}},
 \end{align}
 where $x$ is the distance to the defect.  
 
 In our previous work we found a closed expression for $C_2(\{u_j\})$ valid for any value of $L$ and any value of $M$~\cite{deLeeuw:2015hxa}:
\begin{equation}\label{C-overlap}
 C_2 \left(\left\{u_j\right\}\right) =2\left[
 \left(\frac{2\pi ^2}{\lambda }\right)^L\frac{1}{L}
 \prod_{j}^{}\frac{u_j^2+\frac{1}{4}}{u_j^2}\,\,\frac{\det G^+}{\det G^-}\right]^{\frac{1}{2}},
\end{equation} 
where $G^\pm$ are $\frac{M}{2}\times \frac{M}{2}$ matrices with matrix elements:
\begin{equation}
 G^\pm_{jk}=\left(\frac{L}{u_j^2+\frac{1}{4}}-\sum_{n}^{}K^+_{jn}\right)\delta _{jk}
 +K^\pm_{jk},
\end{equation}
and $K^\pm_{jk}$ are defined as
\begin{equation}
 K^\pm_{jk}=\frac{2}{1+\left(u_j-u_k\right)^2}\pm
 \frac{2}{1+\left(u_j+u_k\right)^2}\, .
\end{equation}

The main result of the present paper is the general formula for the one-point function with arbitrary $k$:
\begin{align}\label{eq:generalOverlap}
C_k \left(\left\{u_j\right\}\right) = 
2^{L-1} C_2\left(\left\{u_j\right\}\right) 
\sum_{j=\frac{1-k}{2}}^{\frac{k-1}{2}}  j^L \prod_{i=1}^{\frac{M}{2}} 
\frac{u_i^2\left(u_i^2 + \frac{k^2}{4}\right)}{\left[u_i^2+(j-\half)^2\right]
\left[u_i^2+(j+\half)^2\right]}\,. 
\end{align}
The multiplicative factor which  relates $C_{2n}$ to $C_2$ is simply the eigenvalue of a product of transfer
matrices of the Heisenberg spin chain when acting on the Bethe state in question and  $C_{2n+1}$ is related to
$C_3$ in a similar manner.  Finally $C_3$ is related to $C_2$ via the eigenvalues of a ratio of $Q$-operators. 
Apart from being deeply rooted in integrability the formula~(\ref{eq:generalOverlap}) also has the appealing property that it
allows us to take a classical, thermodynamical limit which involves scaling $u_i$ in the same way as $k$. An interesting semi-classical limit  with  $k\rightarrow \infty$, $\lambda\rightarrow \infty$ and $\lambda/k^2$ finite exists and allows for a 
comparison of string and gauge theory results. So far, in this limit a match has been found between one-point functions 
of chiral primaries on the string and the gauge theory side~\cite{Nagasaki:2012re,Kristjansen:2012tn}.
 Formula~(\ref{eq:generalOverlap}) opens the possibility
of extending the comparison to massive string states.

The outline of our paper is as follows. 
In section~\ref{AdS/dCFT} we describe in slightly more detail the D3-D5 probe brane set-up and, in  addition,
recapitulate why matrix product states constitute a convenient tool for the calculation of one-point functions. 
Section~\ref{kequal2} contains  some additional insights on the $k=2$ case. Subsequently, in section~\ref{generalk} we proceed to
prove the multiplicative relation between $C_2$ and $C_{2n}$ as well as between
 $C_3$ and $C_{2n+1}$ for $n\geq 2$. Details are relegated to an appendix.  The special case $k=3$ is treated in 
 section~\ref{kequal3}. In section~\ref{large-k} we consider the behavior at large-$k$ and in the thermodynamical
 limit. The latter limit, in principle, allows for a comparison with string theory and
 in section~\ref{stringtheory} we revisit calculation of the one-point functions of the chiral primary states, now from the classical string theory perspective. This set-up bears promise of an  extension to massive states. Finally, section~\ref{conclusion}
 contains some concluding remarks.

\section{One point functions from matrix product states. \label{AdS/dCFT}}

As mentioned above, AdS/CFT set-ups relating probe brane systems with fluxes to defect conformal field theories allow for non-trivial one-point functions. In the simplest such set-up, the D3-D5-brane system, the D5-brane has the geometry $AdS_4\times S^2$ and carries $k$ units of magnetic flux on $S^2$ \cite{Karch:2000gx}. 
On the field theory side one finds ${\cal N}=4$ SYM with a co-dimension one defect separating a region,  $x>0$, where the gauge group is $SU(N)$ from one where it is  $SU(N-k)$ \cite{deWolfe:2001pq,Erdmenger:2002ex}. For $x>0$ the classical equations of motion then allow for a  non-trivial $x$-dependence for some of the
scalar fields, namely~\cite{Constable:1999ac}
 \begin{equation}\label{Phiclass}
 \Phi _i^{\rm cl}=\frac{1}{x}\,\begin{pmatrix}
  \left(t_i\right)_{k\times k} & 0_{k\times (N-k)} \\ 
  0_{(N-k)\times k} & 0_{(N-k)\times (N-k)} \\ 
 \end{pmatrix},~i=1,2,3,\qquad \Phi ^{\rm cl}_i=0,~ i=4,5,6,
\end{equation}
where the three $k\times k$ matrices $t_i$ constitute a unitary $k$-dimensional representation of $\mathfrak{su}(2)$, that is, they satisfy
\begin{equation}\label{eq:su2relations}
 \left[t_i,t_j\right]=i\varepsilon _{ijk}t_k.
\end{equation}
The remaining bulk fields can consistently be set to zero at the classical level. Hence, at tree level the only operators with non-trivial one-point functions (discarding derivatives) are those
which take the form  
\begin{equation}\label{operator}
 \mathcal{O}=\Psi ^{i_1\ldots i_L}\mathop{\mathrm{tr}}\Phi _{i_1}\ldots \Phi _{i_L},
\end{equation}
with $i_1,\ldots, i_L\in \{1,2,3\}$ and, obviously, these one-point functions are obtained simply by replacing each field with its classical value, i.e.
\begin{equation}\label{transformation}
\Psi ^{i_1\ldots i_L}\mathop{\mathrm{tr}}\Phi _{i_1}\ldots \Phi _{i_L}\longrightarrow \Psi ^{i_1\ldots i_L}\mathop{\mathrm{tr}}t _{i_1}\ldots t _{i_L}.
\end{equation}
 The natural basis of operators consists of the operators  with well-defined conformal dimensions and for simplicity
we will restrict our analysis to operators from an $SU(2)$ sub-sector,  a sub-sector known to be closed to all loop orders. 
We therefore define
\begin{eqnarray}\label{whatZ}
 Z=\Phi _1+i\Phi _4,&
\nonumber \\
W=\Phi _2+i\Phi _5,
\end{eqnarray}
and consider only single trace operators built from these two complex scalar fields. Aiming only at tree-level one-point functions it suffices to know
the conformal operators of the theory to one-loop order. It is well-known that the conformal operators in the  $SU(2)$ sub-sector of ${\cal N}=4$ SYM  at one-loop order can be 
identified with the zero-momentum Bethe eigenstates of the XXX$_{1/2}$ Heisenberg spin chain upon mapping each $Z$-field
to a spin up and each $W$-field to a spin down~\cite{Minahan:2002ve}.  This result is unchanged by the presence of the
defect~\cite{DeWolfe:2004zt}. Working within the approach of the algebraic Bethe ansatz
the eigenstates of the XXX$_{1/2}$ Heisenberg spin chain can be written as a series of creation operators
acting on the ferromagnetic vacuum (that we can take to be the state with all spins up), i.e.
\begin{equation}
 \left|\left\{u_j\right\}\right\rangle=B(u_1)\ldots B(u_M)\left|0\right\rangle_L,\label{Bethestate}
\end{equation}
where $L$ denotes the length of the chain, the operator $B(u)$  creates an excitation (a flipped spin)
of rapidity $u$ and in order for the state to be an eigenstate  the rapidities 
$\{u_j\}$ have to fulfill a set of Bethe equations, see for instance \cite{Faddeev:1996iy}. The state~(\ref{Bethestate}) has a total  of $L$ spins and $M$ of these
are down-spins.  It maps to an $SU(2)$ operator built of $L-M$ fields of type $Z$ and $M$ fields of type $W$.

As pointed out in~\cite{deLeeuw:2015hxa} one can implement the transformation (\ref{transformation}) for a given
Bethe eigenstate by taking the inner product of the state with a matrix product state,
defined as 
\begin{equation}\label{MPS}
\left\langle {\rm MPS_k}\,\right|=\mathop{\mathrm{tr}}\nolimits_a
 \prod_{l=1}^{L}\left(\left\langle \uparrow_l\right|\otimes t_1^{(k)}
 +\left\langle \downarrow_l\right|\otimes t_2^{(k)}\right),
\end{equation}
where the index $a$ is an auxiliary space index associated with  the $t_i$'s (and thus takes $k$ different values for a representation of dimension $k$).  Choosing the canonical normalization of the field
theory two-point functions (from the theory without the defect) one can hence express the desired one-point functions as
\begin{equation}\label{genericCso6}
 C_k\left(\left\{u_j\right\}\right)=
 \left(\frac{8\pi ^2}{\lambda }\right)^{\frac{L}{2}}L^{-\frac{1}{2}}\,
 \frac{\left\langle {\rm MPS_k}\,\right.\!\!\left|\vphantom{{\rm MPS}}\left\{u_j\right\}\right\rangle}{\left\langle \left\{u_j\right\} \right.\!\!\left|\left\{u_j\right\}  \right\rangle^{\frac{1}{2}}}.
\end{equation}
Without reference to the dimension of the representation, $k$, one can show that~\cite{deLeeuw:2015hxa}
\begin{itemize}
\item
$C_k\left(\left\{u_j\right\}\right)$ vanishes unless $L$ and $M$ are both even.
\item
$C_k\left(\left\{u_j\right\}\right)$ vanishes unless $\left\{u_j\right\}=\left\{-u_j\right\}$.
\end{itemize}
The states which fulfill the second criterium are the so-called unpaired states which can also be characterized as states being invariant under spin-chain parity, cf.\ f.inst.\ \cite{Beisert:2003tq}. In particular, we note that the one-point function thus effectively depends only on $M/2$
rapidities.  
\section{The $k=2$ case\label{kequal2}}
The overlap for $k=2$ was found in \cite{deLeeuw:2015hxa} and is given by eq.~(\ref{C-overlap}) in the introduction. It was observed in  \cite{deLeeuw:2015hxa} that  for $M=L/2$ the overlap, up to a simple factor, coincided with the overlap between the Bethe eigenstate and the 
N\'{e}el state, i.e.\ the state with alternating spins which is the ground state of the anti-ferromagnetic Heisenberg spin chain
\begin{equation}
 \left|\neel\right\rangle=\left|\ur\dr\ur\dr\ldots \ur\dr\right\rangle
 +\left|\dr\ur\dr\ur\ldots \dr\ur\right\rangle.
\end{equation}
This fact could be exploited to construct a proof of the formula~(\ref{C-overlap}) for $M=L/2$ as it could be proved that, restricted to the components with half-filling, the matrix product state is cohomologically equivalent to the N\'{e}el state 
\begin{equation}\label{MPS->Neel}
\left|\mps_2\right\rangle\Big|_{M=\frac{L}{2}}
 =\frac{1}{2^L(\ihalf)^{M}} \left|\neel\right\rangle
 +S^- \left|\ldots \right\rangle. 
\end{equation}
Here $S_i$ is the total spin operator, and $S^-$ is its lowering component that flips in turn all the spins in the chain with weight one.  Since Bethe eigenstates are highest-weight:
 \begin{equation}
S^+\left|\left\{u_j\right\}\right\rangle=0,
\end{equation}
the second term in~(\ref{MPS->Neel}) does not contribute to the inner product between the matrix product state and
the Bethe state. In this way the result for the one-point function corresponding to a Bethe eigenstate with half filling followed
from the overlap formula for the N\'{e}el state derived in~\cite{Pozsgay:2009}, see also~\cite{Brockmann:2014a,Brockmann:2014b}.  Away from half-filling the formula~(\ref{C-overlap}) continues
to hold.   
This can  be understood from an earlier result for the overlap between a Bethe eigenstate and
the  $(2m)$-fold raised N\'{e}el state~\cite{Brockmann2014}.\footnote{We thank Stefano Mori for pointing this out to us.}  More precisely, it follows by noting
\begin{equation}\label{MPS->Neelgeneral}
 \left|\mps_2\right\rangle \Big|_{M=\frac{L}{2}-2m} = \frac{1}{2^L(\ihalf)^{M}} \frac{1}{(2m)!}({S}^{+})^{2m}\left|\neel\right\rangle
 +S^- \left|\ldots \right\rangle.
\end{equation}
This result directly follows from the fact that the $(2m)$-fold raised N\'{e}el state is equivalent to the generalized N\'{e}el state (compare eq.\ (5.5) from \cite{deLeeuw:2015hxa} and eq.\ (38) from \cite{Brockmann2014}), which was shown to be cohomologically equivalent to the matrix product state in \cite{deLeeuw:2015hxa}. See \cite{FodaZarembo} for a rederivation
of the $k=2$ overlap formula away from half-filling 
using reflecting-boundary domain-wall boundary conditions.

\section{The general $k$ case%
\label{generalk}
}

As explained in~\cite{deLeeuw:2015hxa} one can explicitly evaluate the overlap~(\ref{genericCso6}) for lower values of $L$, $M$ and $k$ by choosing a specific $k$-dimensional representation of $\mathfrak{su}(2)$ and making use of the well-known coordinate space version of the Bethe eigenstates. It was the results of  such evaluations  that first lead us to  the main
result~\eqref{eq:generalOverlap}.

In this section we will prove a recursive relation between 
matrix product states with different values of $k$. More precisely,  we will shown that all matrix product states with  even $k$ are
recursively related  to the matrix product state with $k=2$ via the action of a series of transfer matrices of the Heisenberg
spin chain. Similarly,  all matrix product states with odd $k$ are shown to be recursively related to the matrix product state with $k=3$, and finally evidence is presented that the matrix product state for $k=3$ can be generated from the matrix product state for $k=2$ by the action of a 
ratio of Baxter's $Q$-operators. The general result  \eqref{eq:generalOverlap} then follows from the fact that the Bethe eigenstates are eigenstates of the transfer matrix as well as of Baxter's Q-operator with known eigenvalues.  

For illustrative purposes, let
us spell out the general formula~\eqref{eq:generalOverlap} in a few cases
\begin{align}\label{C3}
C_3 \left(\left\{u_j\right\}\right) &=
\,C_2\left(\left\{u_j\right\}\right) \, 2^L
 \prod_{i=1}^{\frac{M}{2}} 
\frac{u_i^2}{u_i^2 + \frac{1}{4}}\,, \\
C_4 \left(\left\{u_j\right\}\right) &=
 \,C_2\left(\left\{u_j\right\}\right) \,
\left[ 
3^L \prod_{i=1}^{\frac{M}{2}}\frac{u_i^2}{u_i^2 + 1}+ \prod_{i=1}^{\frac{M}{2}}\frac{u_i^2+4}{u_i^2 + 1}\,
\right],
\\
C_5 \left(\left\{u_j\right\}\right) &=
\,C_3\left(\left\{u_j\right\}\right) 
\left[2^L \prod_{i=1}^{\frac{M}{2}}
\frac{u_i^2+\frac{1}{4}}{u_i^2 + \frac{9}{4}}+ \prod_{i=1}^{\frac{M}{2}}\frac{u_i^2+\frac{25}{4}}{u_i^2 + \frac{9}{4}}\right]\,.
\end{align}
The previously announced recursive relation between matrix product states with different values of $k$ takes the following form
\begin{align}\label{eq:recurMPS}
|\mathrm{MPS_{k+2}}\rangle = T(\sfrac{ik}{2}) \, |\mathrm{MPS_k}\rangle - \left(\frac{k+1}{k-1}\right)^L |\mathrm{MPS_{k-2}}\rangle,
\end{align}
where $k\geq2$ and $|\mathrm{MPS}_0\rangle=0$. Here $T(v)$ is the transfer matrix of the XXX$_{1/2}$ Heisenberg spin chain
\begin{align}
T(v)  := \mathrm{tr}_a(\mathcal{L}_{a1} \ldots \mathcal{L}_{aL} ),
\end{align}
with $\mathcal{L}$ the Lax operator
\begin{align}
\mathcal{L}_{a,i}(v) = 1 + \frac{i}{v-\ihalf}P,
\end{align}
which is expressed in terms of the permutation operator $P$.
As usual the label $a$ refers to an auxiliary 2-dimensional space, $\mathbb{C}^2$, which is traced over in the definition of 
$T(v)$. 
For details we refer to~\cite{Faddeev:1996iy}.
The idea behind the proof of formula~(\ref{eq:recurMPS})
is to consider the local action of the Lax operator. 
The matrix product state \eqref{MPS} is an element of $\mathbb{C}^{2L}$ and it is constructed out of the local building blocks
\begin{align}
\Big( \left\langle \uparrow \right|\otimes t_1^{(k)} +\left\langle \downarrow \right|\otimes t_2^{(k)} \Big) \in \mathbb{C}^2 \otimes \mathrm{GL}(\mathbb{C}^k).
\end{align}
Now, we add an additional auxiliary $\mathbb{C}^2$ space and consider the action of the Lax operator on the physical space and the new auxiliary space which gives
\begin{align}\nonumber
\mathcal{L}_{i,a}(\sfrac{ik}{2}) \left[ \left\langle \uparrow_i \right| \otimes t_1^{(k)} +\left\langle \downarrow_i \right|\otimes t_2^{(k)}  \right]  =: 
\left(\left\langle \uparrow_i \right|\otimes \tau_1^{(k)} +\left\langle \downarrow_i \right|\otimes \tau_2^{(k)}\right)
\in \mathbb{C}^2 \otimes \mathrm{GL}(\mathbb{C}^{2k}),
\end{align}
where the matrices $\tau_{1,2}^{(k)}$ are given by
\begin{align}
&\tau_1^{(k)} = 
\begin{pmatrix}
\frac{k+1}{k-1} t^{(k)}_1 & 0 \\
\frac{2}{k-1}t^{(k)}_2 & t^{(k)}_1
\end{pmatrix},
&&\tau_2^{(k)} = 
\begin{pmatrix}
t_2^{(k)} & \frac{2}{k-1}t_1^{(k)} \\
0 & \frac{k+1}{k-1}t_2^{(k)}
\end{pmatrix}.
\end{align}
In the appendix 
 we show explicitly for even as well as for odd $k\geq 2$, that there exists a similarity transformation $A$ such that
\begin{align}
A \tau_i^{(k)} A^{-1} = \begin{pmatrix}
t_i^{(k+2)} & 0 \\
\star_i & \frac{k+1}{k-1}t_i^{(k-2)}
\end{pmatrix}.
\end{align}
This relation immediately proves the recursion relation~(\ref{eq:recurMPS}) for $k\geq 2$. 

The transfer matrix is the key ingredient of the Algebraic Bethe ansatz. In particular, the Bethe states  $|\{u_i\}\rangle$ are eigenvectors of the transfer matrix with eigenvalues
\begin{align}
\Lambda(v|\{u_i\})  =  \left(\frac{v+\ihalf}{v-\ihalf}\right)^L  \prod_{i} \frac{v-u_i-i}{v-u_i} +\prod_{i} \frac{v-u_i+i}{v-u_i}.
\end{align}
The recursion relation \eqref{eq:recurMPS} hence allows us to fix all overlap functions $C_{2n}$ with $n\geq 2$  
in terms of $C_2$ and $C_0\equiv0$, as well as all $C_{2n+1}$ with $n\geq 2$ in terms of $C_3$ and $C_1\equiv0$ 
by means of the following recursion relation
\begin{align}
C_{k+2}  = \Lambda(\sfrac{ik}{2}|\{u_i\}) C_{k} - \left(\frac{k+1}{k-1}\right)^L C_{k-2}.
\end{align}
It is easily checked that \eqref{eq:generalOverlap} (for $k\geq 2$) is the unique solution to this equation.
\section{The special case $k=3$.\label{kequal3}}
The analysis of the previous section involving the transfer matrix did not allow us to prove the relation~(\ref{C3}) for 
$C_3(\{u_j\})$.
However, this relation, which was observed by studying short chains of length $L\leq 10$, seems to indicate that a ratio of
two so-called $Q$-operators could relate $|\mathrm{MPS}_3\rangle$ and $|\mathrm{MPS}_2\rangle$. The $Q$ operator
was originally introduced by Baxter in connection with his solution of the 8-vertex model~\cite{Baxter:1972hz}. Only recently,
an explicit algebraic construction, especially adapted to the $XXX_{1/2}$ Heisenberg chain was carried out~\cite{Bazhanov:2010ts}, see also \cite{Staudacher:2010jz,Bazhanov:2010jq,Frassek:2012mg}. The
Bethe eigenstates are eigenstates of the ${Q}$-operator, i.e.\ they fulfill a relation like
\begin{align}
\hat{Q}(u)  \left|\left\{u_j\right\}\right\rangle \propto \prod_{j=1}^M\left(u-u_j\right) \left|\left\{u_j\right\}\right\rangle.
\end{align}
The algebraic construction of the $Q$-operator from \cite{Bazhanov:2010ts} is strictly speaking only well-defined for the Heisenberg spin chain when a certain twist, $\phi$, is introduced.
The twist can be introduced either at the level of the Hamiltonian or entirely via the boundary conditions. In the latter case
the spin chain boundary conditions turn into
\begin{align}
{\mathcal S}_{L+1}^z={\mathcal S}_1^z,  \hspace{0.5cm} {\mathcal S}_{L+1}^{\pm}=e^{\mp i\phi}{\mathcal S}_1^{\pm}.
\end{align}
 In the presence of the twist, the action of the $Q$ operator on a Bethe eigenstate gives rise to a product of  not
only $M$, but a larger number of factors of the type $(u-u_j)$, hence involving  an extra set of  rapidities which, 
however, all tend to
infinity when the twist is sent to zero. The extra rapidities contain information about Bethe eigenstates in the twisted model which become descendent states in the limit $\phi\rightarrow 0$. Although the $Q$-operator itself is thus ill-defined in the zero twist limit, a ratio of two $Q$-operators
is generically finite and can give rise to exactly the pre-factor in~(\ref{C3}). 

In analogy with the transfer matrix, the $Q$-operator can be defined as the trace of
a certain monodromy matrix~\cite{Bazhanov:2010ts}. The auxiliary Hilbert space associated with the monodromy is infinite dimensional, namely the
Fock space, $\cal{F}$, associated with the usual harmonic oscillator algebra
\begin{align}
\left[\bf{a},\bf{a}^{\dagger}\right]=1.
\end{align}
In other words the auxiliary Hilbert space $\cal{F}$ is spanned by the vectors $|n\rangle$, $n\in Z_0$ which fullfil
\begin{align}
{\bf{a}}^{\dagger} |n\rangle= |n+1\rangle, \hspace{0.5cm} {\bf{a}}  |n\rangle=n  |n-1\rangle.
\end{align}
The $Q$ operator itself then takes the form
\begin{align}
Q(u)  := \frac{e^{\frac{\phi}{2} u}}{\mathrm{tr}_{\cal{F}}\,(e^{-i\phi {\bf{h}}})}
\mathrm{tr}_{\cal{F}}\,\left(e^{-i\phi {\bf{h}}}\,{L}_{L}(u)\otimes \ldots\otimes {L}_{1}(u) \right),
\end{align}
where $\phi$ is the twist,  ${\bf{h}}={\bf{a}}^{\dagger}{\bf{a}}+\frac{1}{2}$, and
\begin{align}
{L}_{l}(u) = \begin{pmatrix}
1 & {\bf{a}}^{\dagger} \\
-i\, {\bf{a}} & u-i {\bf h}
\end{pmatrix}_l.
\end{align}
This explicit form of the $Q$ operator makes it straightforward to implement it in Mathematica and by explicit
computations one can demonstrate that for short matrix product states ($L\leq8$) one has
\begin{align}\label{eq:Q23}
\lim_{\phi\rightarrow 0} \, Q(\sfrac{i}{2})^{-1}Q(0)\,  |\mathrm{MPS}_2\rangle=  2^{-L}\, |\mathrm{MPS}_3\rangle +  S^- |\ldots\rangle.
\end{align}
Note that $ Q(\sfrac{i}{2})^{-1}Q(0)$ is divergent in the $\phi\rightarrow 0$ limit due to the fact that $u=\frac{i}{2}$ corresponds to a singular point for the Bethe equations for any $L$. However, it turns out that the divergencies in the vector $ Q(\sfrac{i}{2})^{-1}\, Q(0) |\mathrm{MPS}_2\rangle$ appear in prefactors of terms of the type  $S^- |\ldots\rangle$ which have zero overlap with a Bethe eigenstate.\footnote{ Thus, strictly speaking
 in order to have a well-defined version of eqn.~(\ref{eq:Q23}), one would have to redefine the left hand side with a term proportional to $S^-$. This is a further complication of the $\phi\rightarrow0$ limit. } We would also like to note that the term $S^- |\ldots\rangle$ first appears for $L=8,M=4$. In particular, for the other values of $L,M$ that were checked, the left hand side of \eqref{eq:Q23} is finite and the ratio of Q-operators exactly relates the matrix product states.\footnote{
One can also consider the equation
\begin{align}\label{eq:Q23b}
Q(0) \,  |\mathrm{MPS}_2\rangle=  2^{-L} Q(\sfrac{i}{2})\, |\mathrm{MPS}_3\rangle + S^- \left|\ldots \right\rangle,
\end{align}
which we found to hold for $L\leq10$ and \textit{any} value of $\phi$. This equation is also divergent in the limit of vanishing twist and again the divergencies are of the form $S^- \left|\ldots \right\rangle$.
}

If formula \eqref{eq:Q23} could be proved true for any length it would immediately imply relation~(\ref{C3}) as the Bethe eigenstates
of the untwisted Heisenberg spin chain are highest weight states. The construction of the $Q$ operator as a monodromy 
matrix makes it tempting to speculate about  the possibility of a proof relying only on the local operator $L_l(u)$, similar in
idea to the proof of the recursive structure of the overlap formula. However, the need for an inversion of $Q$, a limiting procedure as well as the appearance of the term involving the lowering operator complicate matters.

\section{Large $k$ \label{large-k}\label{classical}}
The general formula for the one-point function (\ref{eq:generalOverlap}) is valid for any $k$ under the assumption  that $k\ll N$. An interesting limit to consider is to take $k $ very large (but still small compared to $N$). At strong coupling,  $k$ quite naturally scales with $\lambda $ such that the ratio $k/\sqrt{\lambda }$ remains finite at $\lambda \rightarrow \infty $. This ratio controls the field strength of the internal gauge field on the world-volume of the D5-brane, the holographic dual of the domain wall that separates the two vacua. The classical solution for the D5-brane  \cite{Karch:2000gx} depends only on $k/\sqrt{\lambda }$, but not on $\lambda $ or $k$ separately.

In this paper we study the weak-coupling regime when scaling $k$ with $\lambda $ makes little sense, but we can still take $k\gg 1$. The large-$k$ limit of the overlap that involves a small number of excitations (up to $M=4$) has already been considered in \cite{deLeeuw:2015hxa}. With the explicit expression at hand, we can now take the large-$k$ limit in full generality, for any $M$. We can also consider the thermodynamic limit when the length of the spin chain $L$, the number of excitations $M$ and the rank of the $\mathfrak{su}(2)$ representation $k$ go to infinity simultaneously such that $L\sim M\sim k$.

When  $k$ is large, while $L$ and $M$ are of order one, the sum over $j$ in the general formula \eqref{eq:generalOverlap} is saturated on the upper (or lower) limit of summation:
\begin{align}
C_k \left(\left\{u_j\right\}\right) \simeq 
2^{L-M}C_2\left(\left\{u_j\right\}\right)  \prod_{i=1}^{\frac{M}{2}} u_i^2 
\sum_{j=1}^{\frac{k}{2}} j^{L-2M},
\end{align}
and yields the following result
\begin{align}
C_k \left(\left\{u_j\right\}\right) \simeq \frac{2^{M-1} \prod_{i=1}^{\frac{M}{2}} u_i^2 }{L-2M+1}
C_2\left(\left\{u_j\right\}\right)\,k^{L-M+1}  + \mathcal{O}(k^{L-M}),
\end{align}
whose dependence on $k$ agrees with the scaling indicated in~\cite{deLeeuw:2015hxa} and reproduces in detail the particular cases $M=0,2,4$ studied there.

Alternatively, we can take $k$ to infinity simultaneously with $L$ and $M$. The limit when the spin chain becomes infinitely long and is populated by a large number  of low-lying excitations is the semiclassical limit of the Heisenberg model. The Bethe roots in this regime scale as $u_j\sim L$, while $M\sim L\rightarrow \infty $  
\cite{Sutherland:1995zz,DharShastry,Beisert:2003xu}.
Bethe states of this type describe macroscopic, essentially classical waves of coherent spin precession \cite{Kruczenski:2003gt}. 

While taking the semiclassical limit at weak coupling is not exactly the same as  considering classical strings in $AdS_5\times S^5$, quantities calculated in classical string theory depend on $\lambda $ through the combination $\lambda /L^2$. By re-expanding the string results in this parameter one can often reproduce the weak-coupling perturbation theory up to some fixed order in $\lambda /L^2$. The agreement of the BMN spectrum \cite{Berenstein:2002jq} with magnon energies in the spin chain, or comparison of classical spinning strings in $S^5$ \cite{Frolov:2003qc,Frolov:2003xy} with semiclassical Bethe states  \cite{Beisert:2003xu,Kazakov:2004qf} are two well-known examples where this approach works. In the context of the defect CFT, the one-point functions of protected operators with small $L$ and $M=0$ also perfectly agree with the classical supergravity calculation expanded in $\lambda /k^2$ \cite{Nagasaki:2012re,Kristjansen:2012tn}. Keeping in mind possible comparison to semiclassical string theory (rather than supergravity), we will compute one-point functions of non-protected operators with $M\sim L$ in the thermodynamic limit, taking in addition $k\sim L$ at $L\rightarrow \infty $.

The Bethe roots in the thermodynamic limit condense on a number of cuts in the complex plane and can be characterized by a continuous density
\begin{equation}
 \rho (x)=\frac{1 }{L}\,\sum_{j=1}^{\frac{M}{2}}\left(
 \delta \left(x-\frac{u_j}{L}\right)+\delta \left(x+\frac{u_j}{L}\right)\right).
\end{equation}
The density satisfies a singular integral equation, as a consequence of the Bethe equations for $u_j$'s:
\begin{equation}\label{finite-gap}
 2\strokedint_C\frac{dy\,\rho (y)}{x-y}=\frac{1}{x}+2\pi n_i, \qquad 
 x\in C_i.
\end{equation}
Each of the cuts $C_i$ is associated with an integer mode number $n_i$. The normalization of the density is the filling fraction,
\begin{equation}
 \int_{C}^{}dx\,\rho (x)=\frac{M}{L}\equiv \alpha ,
\end{equation}
and  $\alpha \leq 1/2$ for physical, highest-weight Bethe states.

The ratio of determinants in (\ref{C-overlap}) tends to a constant in the thermodynamic limit  \cite{deLeeuw:2015hxa}, while the products over Bethe roots in (\ref{eq:generalOverlap}) exponentiate and can be replaced by convolution integrals with the density. Approximating summation over $j$ by integration over $\xi=j/L $, we find:
\begin{equation}
 C_k\simeq \,{\rm const}\,\sqrt{L}\left(\frac{8\pi ^2L^2}{\lambda }\right)^{\frac{L}{2}}\int_{-\chi  }^{\chi  }d\xi \,
 \,{\rm e}\,^{LS_{\rm eff}(\xi )},
\end{equation}
where
\begin{equation}
 \chi =\frac{k}{2L}
\end{equation}
and
\begin{equation}
 S_{\rm eff}(\xi )=\frac{1}{2}\int_{}^{}dx\,\rho (x)
 \ln\frac{x^2\left(x^2+\chi  ^2\right)}{\left(x^2+\xi ^2\right)^2}+\ln |\xi|.
\end{equation}

The integral is again saturated at $\xi =\pm \chi $, and we get for the following result for the overlap in the thermodynamic limit:
\begin{equation}
 C_k\sim \frac{\,{\rm const}\,}{\sqrt{L}}\,\left(\frac{2\pi ^2k^2}{\lambda }\right)^{\frac{L}{2}}\,{\rm e}\,^{-\mathcal{A}L},
\end{equation}
where
\begin{equation}\label{curlyA}
 \mathcal{A}=\frac{1}{2}\int_{}^{}dx\,\rho (x)\ln\frac{x^2+\chi  ^2}{x^2}\,.
\end{equation}

The simplest example is the BMN vacuum of the spin chain, the empty state with no Bethe roots that corresponds to the chiral primary operator $\mathop{\mathrm{tr}}Z^L$. In this case $\mathcal{A}=0$, and with exponential accuracy
\begin{equation}\label{weak-l/CPO}
 \left\langle \mathop{\mathrm{tr}}Z^L\right\rangle_{\rm def}\simeq \left(\frac{\sqrt{2}\,\pi k}{\sqrt{\lambda }}\right)^L\frac{1}{R^L}\,,
\end{equation}
where $R$ is the distance from the operator insertion to the defect.

The simplest non-trivial solution of the finite-gap (classical Bethe) equations (\ref{finite-gap}), which is symmetric under $x\rightarrow -x$, has two cuts $(\xx,\yy)$ and $(-\xx,-\yy)$ symmetrically located in the complex plane, such that $\yy=\bar{\mathtt{x}}_1$. The mode numbers of this solution  are $n$ and $-n$. The density can be expressed through elliptic integrals  \cite{Beisert:2003xu}, but it is more convenient to characterize the solution by the quasi-momentum
\begin{equation}
 p(x)=\int_{}^{}\frac{dy\,\rho (y)}{x-y}-\frac{1}{2x}\,,
\end{equation}
whose differential is meromorphic on the two-sheeted cover of the complex plane with cuts. For the two-cut solution \cite{Kazakov:2004qf},
\begin{equation}\label{twocutp}
 dp=\frac{\frac{1}{2}-\alpha -\frac{\xx\yy}{2x^2}}{\sqrt{\left(x^2-\xx^2\right)\left(x^2-\yy^2\right)}}\,dx.
\end{equation}
The solution is parameterized by a single complex variable
\begin{equation}
 r=\frac{\xx^2}{\yy^2}\,,
\end{equation}
through which the endpoints are expressed as
\begin{equation}
 \xx=\frac{1}{4nK}\,,\qquad \yy=\frac{1}{4n\sqrt{r}\,K}\,,
\end{equation}
while the filling fraction is given by
\begin{equation}
 \alpha=\frac{1}{2}-\frac{E}{2\sqrt{r}\,K} \,,
\end{equation}
where $E\equiv E(1-r)$ and $K\equiv K(1-r)$ are the complete elliptic integrals of the second and first kind.

To compute (\ref{curlyA}) we first express its derivative with respect to $\chi $ through the quasi-momentum:
\begin{equation}
 \frac{\partial \mathcal{A}}{\partial \chi }=i\int_{-i\chi }^{i\chi }dp(x)+\frac{1}{\chi }\,.
\end{equation}
Integrating (\ref{twocutp}) twice we obtain:
\begin{eqnarray}
 \mathcal{A}&=&\frac{\chi }{\xx K}\left(EF\left(\varphi \right)-KE\left(\varphi \right)\right)
 -\left(1-2\alpha \right)\ln\frac{\sqrt{\chi ^2+\xx^2}+\sqrt{\chi ^2+\yy^2}}{\xx+\yy}
\nonumber \\
&&
 + \frac{1}{2}\,\ln\frac{\chi ^2\left(\xx^2+\yy^2\right)+2\xx^2\yy^2+2\xx\yy\sqrt{\left(\chi ^2+\xx^2\right)\left(\chi ^2+\yy^2\right)}}{4\xx^2\yy^2}
\nonumber \\
&&
 -\frac{\xx}{\yy}\,\sqrt{\frac{\chi ^2+\yy^2}{\chi ^2+\xx^2}}+1,
\end{eqnarray}
where $F(\varphi )$ and $E(\varphi )$ are the incomplete elliptic integrals of the same modulus $1-r$, and argument
\begin{equation}
 \tan\varphi =\frac{\chi }{\xx}\,.
\end{equation}

The one-point function exponentiates in the thermodynamic limit, which suggests a semiclassical interpretation. Since the exponent is always negative (it is easy to see that $\mathcal{A}>0$), the overlap is exponentially suppressed and perhaps can be interpreted as a tunneling amplitude of a transition between a Bethe eigenstate and the MPS or generalized N\'eel  state. If this interpretation is correct the transition amplitude could probably be described  in the semiclassical regime by an instanton solution of the Landau-Lifshitz equations, the classical equations of motion of the Heisenberg model. We are not in a position to construct such a solution here. Instead we will study the one-point function at strong coupling, where the description from the very beginning is in classical terms.

\section{Comparison to string theory\label{stringtheory}}

In string theory, the defect that separates the $SU(N)$ and $SU(N-k)$ vacua is described by a D5-brane embedded in $AdS_5\times S^5$, and carrying $k$ units of magnetic flux on its world-volume. The magnetic flux naturally scales with $\lambda $ such that 
\begin{equation}
 \kappa =\frac{\pi k}{\sqrt{\lambda }}
\end{equation}
remains finite in the strong-coupling limit. 

The brane embedding is very simple in  Poincar\'e coordinates
\begin{equation}
 ds^2=\frac{dx^2+dz^2}{z^2}\,.
\end{equation}
The brane intersects $AdS_5$  along the $AdS_4$ hyperplane, tilted with respect to the boundary at an angle that depends on the magnetic flux \cite{Karch:2000gx,Nagasaki:2012re}:
\begin{equation}\label{D5-embedding}
 x=\kappa z,
\end{equation}
where $x$ is the direction perpendicular to the defect (for instance, $x=x^3$ if the defect is the domain wall in the $x^1-x^2$ plane).
The remaining two dimensions of the brane wrap the equatorial two-sphere in $  S^5$. The orientation of the $S^2$ within $S^5$ is dictated by the R-symmetry quantum numbers of the defect. The solution (\ref{Phiclass}) involves scalar fields $\Phi _1$, $\Phi _2$, $\Phi _3$. When $S^5$ is represented by a unit sphere in $\mathbbm{R}^6$, each $\Phi _i$ is dual to the $i$-th coordinate direction, and consequently the D-brane intersects the $S^5$ along  the 
$123$-plane. The background gauge field is a constant (monopole) magnetic field with $k$ units of field strength on $S^2$.

The process of emitting or absorbing a string by a D-brane is described by a string world-sheet attached to the D-brane at its constant-$\tau $ section, which we take to be $\tau =0$. The boundary conditions are then of the Dirichlet type for the coordinates transverse to the brane ($X^i$) and mixed Neumann-Dirichlet for the longitudinal coordinates $X^\mu $:
\begin{eqnarray}\label{DN-BC}
 &&\partial _\sigma  X^i=0,
\nonumber \\
 &&\partial _\tau X_\mu +\frac{2\pi }{\sqrt{\lambda }}\,F_{\mu \nu }\,\partial _\sigma X^\nu =0,
\end{eqnarray}
where $F_{\mu \nu }$ is the internal gauge field on the D-brane world-volume.

The one-point function of a local operator is computed by  inserting a vertex operator in the string path integral:
\begin{equation}
 \left\langle \mathcal{O}(x)\right\rangle_{\rm def}=\int_{}^{}DX^M\,
 \int_{}^{}d^2w\,V_{\mathcal{O}}\left(X(w)|x\right)\,{\rm e}\,^{-\frac{\sqrt{\lambda }}{2\pi }\,S_{\rm str}[X]}.
\end{equation}
The vertex operator, schematically, has the following form:
\begin{equation}
 V(X|x)=\partial X\,\partial X\,\,{\rm e}\,^{\Sigma (X)},
\end{equation}
where, roughly speaking, $\,{\rm e}\,^{\Sigma }$ is the wave function of the corresponding string mode in $AdS_5\times S^5$. The exponent is proportional to the quantum number of the string state,  and for large quantum numbers, $\Sigma \sim Q\sim \sqrt{\lambda }$ is of the same order of magnitude as the string action.

In the semiclassical approximation, valid at $\lambda \rightarrow \infty $, the path  integral over $X^M$, as well as the integration over the position of the vertex operator, are saturated on the saddle point of the integrand:
\begin{equation}
 \left\langle \mathcal{O}(x)\right\rangle_{\rm def}\simeq V_{\mathcal{O}}\left(X_{\rm cl}(w_0)|x\right)\,{\rm e}\,^{-\frac{\sqrt{\lambda }}{2\pi }\,S_{\rm str}[X_{\rm cl}]},
\end{equation}
where $\simeq $ denotes equality with exponential accuracy, and $X^M_{{\rm cl}}$ is the solution of the string equations of motion \cite{PolyakovStrings2002,Tseytlin:2003ac}:
\begin{equation}
 \frac{\delta S_{\rm str}}{\delta X^M}=\frac{2\pi }{\sqrt{\lambda }}\,\,
 \frac{\partial \Sigma }{\partial X^M}\,\delta \left(w-w_0\right).
\end{equation}

The boundary conditions are the Dirichlet-Neumann ones (\ref{DN-BC}) on the end of the string which is attached to the D-brane. The delta-function source in the equations of motion can be traded for boundary conditions at the other end (at $w\rightarrow w_0$). In the simplest case both the equations of motion and the source can be linearized near $w=w_0$:
\begin{eqnarray}
 &&\frac{\delta S_{\rm str}}{\delta X^M}= -\partial ^2X_M+\ldots 
\nonumber \\
&&\Sigma = Q_MX^M+\ldots 
\nonumber 
\end{eqnarray}
The delta-function then produces a logarithmic singularity at $w_0$ in $X^M$:
\begin{equation}
 X_M=-\frac{Q_M}{\sqrt{\lambda }}\,\ln|w-w_0|+\ldots 
\end{equation}
It is convenient to introduce exponential coordinates near $w$:
\begin{equation}
 w-w_0=\,{\rm e}\,^{i\sigma -\tau },
\end{equation}
The boundary conditions then take the familiar form of the string moving in the direction $X^M$ with momentum $Q_M$:
\begin{equation}\label{freemotion}
 X_M=\frac{Q_M}{\sqrt{\lambda }}\,\tau +\ldots 
\end{equation}

In this paper we only consider the simplest case of the chiral primary operator $\mathcal{O}=\mathop{\mathrm{tr}}Z^L$. The dual vertex operator is known exactly \cite{Berenstein:1998ij}, but for our purposes the exponential accuracy would suffice: 
\begin{equation}\label{BMNvertex}
 V_{\rm CPO}\simeq 
 2^\frac{L}{2}z^{-L}\,{\rm e}\,^{iL\varphi }.
 \end{equation}
Here $\varphi $ is the angle in the $14$-plane in $\mathbbm{R}^6$ (the orientation is again dictated by the R-symmetry quantum numbers of the field $Z$ in (\ref{whatZ})).

The classical string solution with the boundary conditions described above can
 be constructed by the method of images, placing a fictitious source at the same distance $R$ on the other side of the defect and considering a two-point function $\left\langle \mathop{\mathrm{tr}}\bar{Z}^L(-R)\mathop{\mathrm{tr}}Z^L(R)\right\rangle$.
  The classical solution for the latter is the Euclidean continuation of the BMN geodesic \cite{Dobashi:2002ar,Tsuji:2006zn,Janik:2010gc}:
\begin{eqnarray}\label{BMN-string}
 \varphi &=&i\omega \tau, 
\nonumber \\
x&=&R\tanh\omega \left(\tau +\tau _0\right),
\nonumber \\
z&=&\frac{R}{\cosh\omega \left(\tau +\tau _0\right)}\,.
\end{eqnarray}
We take the $\tau >0$ portion of the world-sheet as the solution for the string ending on the D-brane.
The solution automatically satisfies the right boundary conditions at the operator insertion point ($\tau =\infty $), provided that
\begin{equation}
 \omega =\frac{L}{\sqrt{\lambda }}\,,
\end{equation}
which follows from comparing (\ref{freemotion}) with (\ref{BMNvertex}).

\begin{figure}[t]
\begin{center}
 \centerline{\includegraphics[width=10cm]{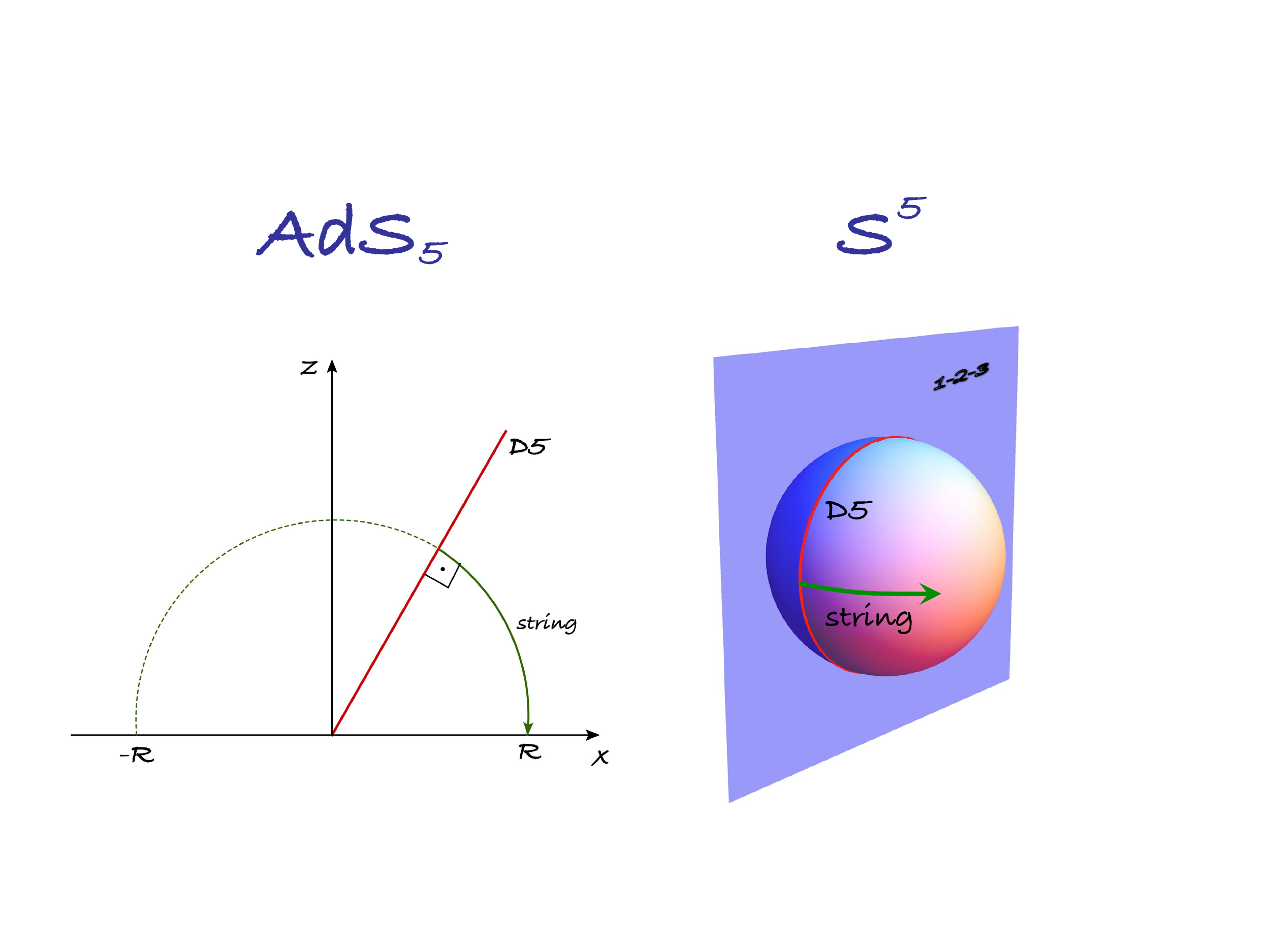}}
\caption{\label{BMN-fig}\small The BMN string ending on the brane can be constructed by the method of images from the solution that describes the two-point function of the operators inserted at points $-R$ and $R$ placed symmetrically on the two sides of the defect. Since the string world-line in $AdS_5$, geometrically, is a semicircle, it is perpendicular to the D5-brane and hence satisfies the correct Dirichlet-Neumann boundary conditions.}
\end{center}
\end{figure}

As for the boundary conditions on the D-brane, the solution can be made compatible with them by adjusting the constant of integration $\tau _0$. For a point-like string the boundary conditions are Dirichlet in the transverse directions and purely Neumann in the longitudinal ones (the magnetic field plays no r\^ole because $\partial _\sigma X^\mu =0$). Geometrically, these boundary conditions mean that the string world-sheet should meet the D-brane (\ref{D5-embedding}) at the right angle. And indeed, the string forms a semi-circle centered at zero, for which  the D5-brane (projected onto the $xz$ plane) is a radius and so the two are perpendicular at the point of intersection (fig.~\ref{BMN-fig}). The same is true on $S^5$, where the string trajectory approaches the brane, sitting in the $123$-plane, perpendicularly, along the $14$-plane.

It is only necessary to make sure that string emission is simultaneous in $S^5$ and $AdS_5$. This can done by adjusting the parameter $\tau _0$.
The condition for $X^M(0)$ to lie on the D-brane worldvolume (\ref{D5-embedding}) is
\begin{equation}\label{tau0}
 \kappa =\frac{x(0)}{z(0)}=\sinh\omega \tau _0,
\end{equation}
which determines $\tau _0$ in terms of $\kappa $.

The relevant part of the string action (the solution is in the conformal gauge),
\begin{equation}
 S=\frac{1}{2}\int_{}^{}d\tau d\sigma \,\left[\frac{\left(\partial x\right)^2+\left(\partial z\right)^2}{z^2}+\left(\partial \varphi \right)^2\right],
\end{equation}
evaluates to zero on the classical solution (\ref{BMN-string}), so the contribution to the one-point functions comes entirely from the vertex operator. From (\ref{BMNvertex}) we find:
\begin{equation}\label{trzL}
  \left\langle \mathop{\mathrm{tr}}Z^L\right\rangle_{\rm def}\simeq 
  \frac{2^\frac{L}{2}}{R^L}\lim_{\tau \rightarrow \infty }\cosh^L\omega \left(\tau +\tau _0\right)\,{\rm e}\,^{-\omega L\tau }=
  \left(\frac{\kappa +\sqrt{\kappa ^2+1}}{\sqrt{2}}\right)^L
  \,\frac{1}{R^L}\,,
\end{equation}
where we have used (\ref{tau0}) in the last equality. This result agrees with the supergravity calculation \cite{Nagasaki:2012re} in their overlapping regime of validity. One  can check that at large\footnote{The R-charge is denoted by $l$ in \cite{Nagasaki:2012re}.} $L$ , the integral (3.15) in \cite{Nagasaki:2012re} is saturated by the saddle-point which results in (\ref{trzL}). This is not surprising, since at large $L$ the geodesic approximation should be valid for the supergraviton propagator in $AdS_5$, making string and supergravity calculations manifestly equivalent.

This result is valid at $\lambda \rightarrow \infty $, $k\rightarrow \infty $, with $k/\sqrt{\lambda }$ fixed. In this approximation the one-point function does not depend on $\lambda $ and $k$ separately but only on their ratio, and we have not made any assumptions on whether this ratio is big or small. Assuming that $\kappa $ is big we can expand the answer in $1/\kappa ^2=\lambda /\pi ^2k^2$ getting a power series that resembles in form the ordinary perturbation theory. To leading order we get: 
\begin{equation}
 \left\langle \mathop{\mathrm{tr}}Z^L\right\rangle_{\rm def}\simeq
 \frac{1}{R^L}\,\left(\frac{\sqrt{2}\pi k}{\sqrt{\lambda } }\right)^L\left(1+O\left(\frac{\lambda }{k^2}\right)\right),
\end{equation}
in complete agreement with the weak-coupling prediction (\ref{weak-l/CPO}).

\section{Conclusions\label{conclusion}}

Our general form for the one-point function~(\ref{eq:generalOverlap}) hints that integrability may play a more profound role in the present
context than we have so far been able to reveal. First, the recursive structure of the one-point function formula
hinges on the properties of the transfer matrix of the Heisenberg spin chain and seems to indicate the possibility of 
a proof which builds more directly on the algebraic Bethe ansatz approach. Secondly, the relation between the results
for the two- and three-dimensional representation points towards a novel application of Baxter's Q-operator.

As pointed out previously, one-point functions of chiral primary operators calculated in  dCFT's have been successfully matched
to one-point functions calculated in a supergravity approach~\cite{Nagasaki:2012re,Kristjansen:2012tn}. 
The fact that we have derived an overlap formula valid for any value of $k$ opens up a vast new arena for the comparison
between field theory and string theory, namely the comparison of one-point functions of massive operators. 
We have taken a first step towards entering this arena by re-formulating the gravity computation of the chiral primary one-point
function in a way which in principle allows for a generalization to massive states. Implementing this generalization constitutes 
an interesting and challenging future line of investigation.

Our work points towards several other possible lines of investigation. One-point functions on the field theory side 
could be studied at higher loop orders, in bigger sectors or for other types of defect field theories resulting from probe-brane
set-ups with fluxes. It would also be interesting to investigate in further detail the 
exact role of the Q-operator in the present context and to find a proof of \eqref{eq:Q23}.

\section*{Acknowledgments}

We would like to thank, R.\ Frassek, A.\ Ipsen, S.\ Mori and
G.W.\ Semenoff for interesting discussions. 
C.K and K.Z. would like to thank G.W.\ Semenoff for kind hospitality at UBC, Vancouver, where part of this work had been done.
I.B.-M.,  M.d.L. and C.K.\ were supported in part by FNU through grants number DFF-1323-00082 and DFF-4002-00037.
The work of K.Z. was supported by the Marie Curie network GATIS of the European Union's FP7 Programme under REA Grant
Agreement No 317089, by the ERC advanced grant No 341222, by the Swedish Research Council (VR) grant
2013-4329, and by RFBR grant 15-01-99504. 

\appendix

\section{Similarity transformation}\label{app:SimTrans}
In this section we present a similarity transformation matrix $A$ and the matrix quantities $\star_i$ which fulfill
\begin{align}
&A \tau_i^{(k)} A^{-1} = \hat t_i^{\,(k)}, && i=1,2,
\end{align}
where
\meq{\hat t_i^{\,(k)} = \begin{pmatrix}
t_i^{(k+2)} & 0 \\
\star_i & \frac{k+1}{k-1}t_i^{(k-2)}
\end{pmatrix}.}

The quantities $A$, $A^{-1}$ and $\star_i$ are expressed in terms of the matrix unities $E^i_j$ for which 
 \meq{E^i_j\, E^k_l = \delta^k_j\,  E^i_l.}
 
It is then a tedious albeit straightforward computation to show that
\meq{AA^{-1} = 1, \qquad \mt{and} \qquad A\tau_i^{(k)} = \hat t_i A.}

\subsection*{Constructing $A$}
We define the following functions
\meq{K[k,j] =  \left(\frac{k+1}{k-1}\right)^{\frac{(j-2) (j+1)}{4}} \left(\frac{k-2}{k}\right)^{\frac{(j-2) (j-1)}{4}},}
\meq{F[k] =\frac{k(k-1)}{k+1}\sqrt{\frac{k-1}{k-2}},}
\meq{H[k] = \frac{\sqrt{2}}{k+1}\sqrt{\frac{k-1}{k-2}}}
and
\meq{G[j] = \sqrt{j(j+1)}.}
Furthermore the matrix structure is such that we can write it in terms of the matrices
\meq{Z^n_m = E^n_m + i\, E^n_{k+m}, \qquad \mt{and} \qquad W = Z^T,}
where $E^n_m$ are the $2k \times 2k$ matrix unities -- the one appears in the $n$-th row in column $m$. We will also use the complex conjugates of these matrices, and we denote them $\bar Z, \bar W$.

\subsubsection*{Even $k$}

The similarity transformation for even values of $k$ is given by
\begin{align}
A_\mt{even} = & \; Z_1^{k+3}-\bar Z_{k}^{2 k} + H[k] \left(Z_{k-2}^{2 k}-\bar{Z}_3^{k+3}\right) \\
 &+ \sum_{j\,=\,1}^{k}\frac{G[j]}{G[k-1]}\left(Z_j^{j+2} + {\bar Z}_{k-j+1}^{k-j+1} \right) +  \sum_{j\,=\,2}^{\floor{ k/2 -1}}\frac{ F[k]}{G[j]}\left(Z_{k-j-1}^{2 k-j+1}-\bar{Z}_{j+2}^{k+j+2}\right) \nonumber
 \end{align}
and its inverse by
\begin{align}
A_\mt{even}^{-1} =& \; \frac{1}{2}\sum _{j\,=\,1}^3K[k,j]\left(W_j^j  + \bar W_{k-j+3}^{k-j+1}\right)  +  \frac{1}{2}\frac{G[1]}{G[k-1]}\left(W_{k}^{k}+\bar W_3^1\right)\\
& + \frac{1}{2} \sum_{j\,=\,1}^{k-2} \frac{G[j]}{F[k]}\left(\bar W_{2 k-j+1}^{k-j-1} - W_{k+j+2}^{j+2}\right) +  \frac{1}{2}\sum_{j\,=\,2}^{\floor{k/2 -1}}\frac{G[k - 1]}{G[j]}\left(W_{k-j+1}^{k-j+1}+\bar W_{j+2}^j\right).\nonumber
\end{align}

\subsubsection*{Odd $k$}
For odd values of $k$ the similarity transformation is given by
\eq{A_\mt{odd} = A_\mt{even}  + \frac{F[k]}{2\, G\left[\frac{k-1}{2}\right]}\left(Z_{\frac{k-1}{2}}^{\frac{3 (k+1)}{2}}-\bar Z_{\frac{k+3}{2}}^{\frac{3 (k+1)}{2}}\right)}
and its inverse by
 \eq{A_\mt{odd}^{-1} = A^{-1}_\mt{even} +  \frac{G[2 k]}{G[2 k+1]}\left(W_{\frac{k+3}{2}}^{\frac{k+3}{2}}+\bar W_{\frac{k+3}{2}}^{\frac{k+3}{2}-2}\right).}
 

\subsection*{Constructing $\star_i$}
We define the following functions
\meq{F^\star[k,j] = k\left(\frac{k-1}{k+1}\right) \sqrt{\frac{k}{k-2}}  \sqrt{\frac{k-2-j}{(j+1)(j+2)(j+3)}},}
\meq{G^\star[k] = (k-1)\left(\frac{k-1}{k+1}\right) \sqrt{\frac{k}{k-2}}  \sqrt{\frac{k+2}{k-2}},} 
\meq{H^\star[k] = \frac{k}{2}\left(\frac{k-1}{k+1}\right)\sqrt{\frac{k}{k-2}}\sqrt{\frac{k+3}{k-1}},}
\meq{I^\star[k] = \frac{k}{2}\sqrt{\frac{k}{k-2}}\sqrt{\frac{k-1}{k-3}},}
and
\meq{J^\star[k] = \frac{k+1}{2} \sqrt{\frac{k}{2}}\sqrt{\frac{k-3}{k-1}}.}
Then $\star_i$  $(i = 1,2)$ is given by
\begin{align}
\star_i = &\, (-1)^{\frac{i-1}{2}} \bigg[ \sum_{j\,=\,1}^{\floor{k/2 -3}}F^\star[k,j] ( E^{2k-j}_{k-j-1} +  (-1)^i E^{k+j+3}_{j+4})  \\
& - J^\star[k](E^{k+4}_3 + (-1)^i E^{2k-1}_{k} )- \frac{k+3}{2k}F^\star[k,0]\left(E^{2k}_{k-1} +  (-1)^i E^{k+3}_4 \right) +\mc R^\star_i[k]\bigg]\nonumber
\end{align}
where
\meq{\mc R^\star_i[k] =\! \left\{\begin{array}{l c}G^\star[k](E^{\frac{3k+4}{2}}_{\frac{k+2}{2}} + (-1)^i E^{\frac{3k+2}{2}}_{\frac{k+4}{2}} ), & \mt{ even } k\\[3mm]
I^\star[k](E^{\frac{3k+5}{2}}_{\frac{k+3}{2}} + (-1)^i E^{\frac{3k+1}{2}}_{\frac{k+3}{2}}) + H^\star[k](E^{\frac{3k+3}{2}}_{\frac{k+1}{2}} + (-1)^i E^{\frac{3k+3}{2}}_{\frac{k+5}{2}}), & \mt{ odd } k \end{array}\right.}
Note that we have written $\star_i$ as a $2k \times 2k$ dimensional matrix. To be clear this is just to ease the computation, and strictly speaking $\star_i$ denotes the $(k-2)\times(k+2)$ matrix sitting inside $\hat t_i^{(k)}$. It's simply a matter of taking $E^n_m \rightarrow \check E^{n-k-2}_m$ where the latter is a matrix unity of dimension $(k-2)\times(k+2)$.

\bibliographystyle{nb}
\bibliography{refs}

\end{document}